\documentclass[final]{rjparticle}
\usepackage{graphicx}

\DeclareSymbolFont{extraup}{U}{zavm}{m}{n}
\DeclareMathSymbol{\varheart}{\mathalpha}{extraup}{86}
\DeclareMathSymbol{\vardiamond}{\mathalpha}{extraup}{87}

\newcommand{\cinf}{{{\cal \cC}^\infty(M,\R)}}

\def\ad{\mathrm{ad}}
\newcommand{\D}{{\check{D}^\ad}}

\newcommand{\cone}{\mathrm{cone}}

\newcommand{\bp}{\begin{Proposition}}
\newcommand{\ep}{\end{Proposition}}
\newcommand{\bl}{\begin{Lemma}}
\newcommand{\el}{\end{Lemma}}
\newcommand{\bt}{\begin{Theorem}}
\newcommand{\et}{\end{Theorem}}
\newcommand{\bd}{\begin{Definition}}
\newcommand{\ed}{\end{Definition}}
\newcommand{\End}{\mathrm{End}}

\newcommand{\ev}{\mathrm{ev}}
\newcommand{\eqdef}{\stackrel{{\rm def.}}{=}}

\DeclareFontFamily{U}{rsf}{}
\DeclareFontShape{U}{rsf}{m}{n}{<5> <6> rsfs5 <7> <8> <9> rsfs7 <10-> rsfs10}{}
\DeclareMathAlphabet\Scr{U}{rsf}{m}{n}

\newcommand{\KA}{K\"{a}hler-Atiyah~}

\def\Z{{\mathbb Z}}

\def\R{{\mathbb R}}

\def\dd{\mathrm{d}}

\def\AdS{\mathrm{AdS}}

\newcommand{\be}{\begin{equation*}}
\newcommand{\ee}{\end{equation*}}
\newcommand{\ben}{\begin{equation}}
\newcommand{\een}{\end{equation}}
\newcommand{\beqa}{\begin{eqnarray*}}
\newcommand{\eeqa}{\end{eqnarray*}}
\newcommand{\beqan}{\begin{eqnarray}}
\newcommand{\eeqan}{\end{eqnarray}}

\newcommand{\nn}{\nonumber}

\renewcommand{\id}{\mathrm{id}}

\def\cC{{\mathcal C}}

\def\cB{\Scr B}

\def\Cl{\mathrm{Cl}}

\def\cC{\mathcal{C}}

\def\G_2{\mathrm{G_2}}

\newcommand{\twopartdef}[4]
{
	\left\{
		\begin{array}{ll}
			#1 & \mbox{if } #2 \\
			#3 & \mbox{if } #4
		\end{array}
	\right.
}

\title{Geometric algebra and M--theory compactifications} 
\author[1]{Calin Iuliu Lazaroiu}
\author[1]{Elena-Mirela Babalic}
\affil[1]{Department of Theoretical Physics,\\
``Horia Hulubei'' National Institute for Physics and Nuclear Engineering,\\
Reactorului 30, RO-077125, POB-MG6, M\u{a}gurele-Bucharest, Romania\\
{\em Email}: lcalin@theory.nipne.ro, mbabalic@theory.nipne.ro}

\begin{document}
\maketitle
\begin{abstract}
We show how supersymmetry conditions for flux compactifications of
supergravity and string theory can be described in terms of a flat subalgebra
of the \KA algebra of the compactification space, a description which has
wide-ranging applications.  As a motivating example, we consider the most
general M-theory compactifications on eight-manifolds down to $\AdS_3$ spaces
which preserve $N=2$ supersymmetry in 3 dimensions.  We also give a brief sketch of the lift of
such equations to the cone over the compactification space and of the
geometric algebra approach to `constrained generalized Killing pinors', which
forms the technical and conceptual core of our investigation.
\end{abstract} 

\section{Introduction}

Despite their physical relevance, the most general $N=2$ warped product 
compactifications of 11-dimensional supergravity down to a three-dimensional Anti-de-Sitter 
space have not previously been studied in detail. In such compactifications, the internal eight-manifold $M$
carries a Riemannian metric $g$ as well as a one-form $f$ and a four-form $F$, the latter two 
of which encode the 4-form field strength of the eleven-dimensional theory. The internal Majorana pinor
is a section of the real pin bundle $S$ of $M$, which is a real vector bundle of rank $16$. 
The condition that such a background preserves exactly $N=2$ supersymmetry in 3 dimensions amounts 
to the requirement that the real vector space of solutions to the following algebro-differential 
system (the so-called {\em constrained generalized Killing  (CGK) pinor equations}): 
\ben
\label{par_eq}
D\xi = Q\xi=0~~
\een
has dimension two. Here, $Q$ is an endomorphism of $S$ given by:
\be
Q=\frac{1}{2}\gamma^m\partial_m\Delta -\frac{1}{288}F_{m p q r}\gamma^{m p q r}
-\frac{1}{6}f_p \gamma^p \gamma^{(9)}
-\kappa\gamma^{(9)} ~~
\ee
while $D=\nabla^S+A$ is a linear connection on $S$, where $\nabla^S$ is the
connection induced on $S$ by the Levi-Civita connection of $(M,g)$ and $A=\dd
x^m\otimes A_m\in \Omega^1(M,\End(S))$ is an $\End(S)$-valued one form on $M$,
with:
\be
A_m= \frac{1}{4}f_p\gamma_{m}{}^{p}\gamma^{(9)}+\frac{1}{24}F_{m p q r}\gamma^{ p q r}+\kappa \gamma_m\gamma^{(9)}\in \Gamma(M, \End(S))~~.
\ee
The quantity $\kappa$ is a positive parameter related to the cosmological
constant $\Lambda$ of the $\AdS_3$ space through $\Lambda=-8\kappa^2$.  
The requirement of having $N=2$ supersymmetry in three-dimensions does not
impose any chirality condition on $\xi$. If one imposes such a condition as an
{\em extraneous} technical assumption (for example, if one adds the condition
$\gamma^{(9)}\xi=+\xi$ to the system \eqref{par_eq}), then one obtains a
drastic simplification leading to the well-known results of \cite{Becker}. For
a number of reasons (having to do, in particular, with efforts to generalize
F-theory) we are interested in studying the problem without imposing any such
chirality constraint. This leads to unexpected complications, which can be
resolved upon re-formulating the problem by using geometric algebra
techniques.

\section{The geometric algebra approach to pinors}

The standard construction of the pin bundle $S$ of a (pseudo)-Riemannian
manifold $(M,g)$ of signature $(p,q)$ and dimension $d=p+q$ can be described most briefly by saying that
$S$ is a bundle of modules over the Clifford bundle $\Cl(T^\ast M)$ of the
cotangent bundle of $M$ --- where $T^\ast M$ is, of course, endowed with the
metric ${\hat g}$ induced by $g$ \footnote{This is equivalent \cite{Trautman} with giving $S$ as
the vector bundle associated with a ${\rm Clifford}^c$-structure of $M$ via a
representation of the ${\rm Clifford}^c$ group.}.  One problem with this
approach (which manifests itself in many subtle aspects of spin
geometry as constructed in \cite{SpinGeometry}) is that the Clifford bundle
is determined by $(M,g)$ only up to isomorphism and hence the association of
$\Cl(T^\ast M)$ to $(M,g)$ is {\em not} functorial. The issue can be resolved
by using a particular realization of the Clifford bundle (going back to Chevalley \cite{Chevalley} and Riesz \cite{Riesz}) which is known as the \KA
bundle of $(M,g)$. This removes the ambiguities of
the standard approach to spin geometry, since the \KA bundle of $(M,g)$ is
functorially determined by $(M,g)$. The Chevalley-Riesz realization
identifies the underlying vector bundle of $\Cl(T^\ast M)$ with the exterior
bundle $\wedge T^\ast M$ of $M$, transporting the Clifford product of the
former to a non-commutative but unital and associative fiberwise
multiplication on the latter which we denote by $\diamond$ and call the {\em
geometric product} of $(M,g)$. The geometric product makes $\wedge T^\ast M$
into the {\em \KA bundle} $(\wedge T^\ast M, \diamond)$, a bundle of associative algebras which is {\em naturally}
(i.e. functorially) determined by $(M,g)$. The geometric product (which
depends on $g$) is not homogeneous with respect to the natural $\Z$-grading
(given by rank) of the exterior bundle. However, it admits an expansion into a finite sum
of binary operations $\bigtriangleup_k$ ($k=0\ldots d$) which are homogeneous
of degree $-2k$ with respect to that grading:
\ben
\label{starprod}
\begin{split}
\diamond =\sum_{k=0}^{\left[\frac{d}{2}\right]}
(-1)^k \bigtriangleup_{2k} + \sum_{k=0}^{\left[\frac{d-1}{2}\right]} (-1)^{k+1} \bigtriangleup_{2k+1}\circ (\pi\otimes \id_{\wedge T^\ast M})~~,
\end{split}
\een
where $\pi$ is the parity automorphism, which is defined through: 
\be \label{parity}
\pi\eqdef \oplus_{k=0}^d(-1)^k\id_{\wedge^k T^\ast M}~~.
\ee
The binary products $\bigtriangleup_k:\wedge T^\ast M\times_M \wedge T^\ast M\rightarrow T^\ast M$ are known as {\em generalized products}. Expansion 
\eqref{starprod} can be viewed as the semiclassical expansion of the geometric product when the latter is identified with the star product arising in 
a certain `vertical' geometric quantization procedure in which the role of the Planck constant is played by the inverse of the overall 
scale of the metric $g$. In particular, the classical limit corresponds to $g\rightarrow \infty$ and $\diamond$ reduces to the wedge product 
$\wedge=\bigtriangleup_0$ in that limit. The other generalized products $\bigtriangleup_k$ ($k>0$) depend on $g$, being determined on sections 
of $\wedge T^\ast M$ by the recursion formula: 
\be
\begin{split}
\omega\bigtriangleup_{k+1}\eta
=\frac{1}{k+1}g^{ab}(e_a \lrcorner \omega)\bigtriangleup_{k}(e_b\lrcorner \eta)=g_{ab}
(\iota_{e^a}\omega) \bigtriangleup_k (\iota_{e^b}\eta) ~~,
\end{split}
\ee
where $\iota$ denotes the so-called {\em interior product} \cite{ga1}. 

The pin bundle $S$ can now be viewed as a bundle of modules over the \KA
bundle of $(M,g)$, where the module structure is defined by a morphism of
bundle of algebras which we denote by $\gamma:(\wedge T^\ast
M,\diamond)\rightarrow (\End(S),\circ)$. Since we are interested in pinors of
spin $1/2$, we assume that $\gamma$ is fiberwise irreducible. 

\paragraph{Notations and conventions.} 

We let $(e_m)_{m=1\ldots 8}$ denote a local frame of $T M$, defined on some
open subset $U\subset M$ and $e^m$ be the dual local coframe (local frame of
$T^\ast M$), which satisfies $e^m(e_n)=\delta^m_n$ and ${\hat
g}(e^m,e^n)=g^{mn}$, where $(g^{mn})$ is the inverse of the matrix $(g_{mn})$.
The space of smooth inhomogeneous globally-defined differential forms on $M$
is denoted by $\Omega(M)\eqdef \Gamma(M,\wedge T^\ast M)$. The fixed rank
components of the graded module $\Omega(M)$ are denoted by
$\Omega^k(M)=\Gamma(M,\wedge^k T^\ast M)$, with $k=0,\ldots, \dim M$.  A
general inhomogeneous form $\omega\in \Omega(M)$ expands as:
\ben
\label{FormExpansion}
\omega=\sum_{k=0}^d\omega^{(k)}=_{U}\sum_{k=0}^{d}\frac{1}{k!}\omega^{(k)}_{a_1\ldots
a_k} e^{a_1\ldots a_k}~~{\rm with}~~\omega^{(k)}\in \Omega^k(M)~~,
\een
where $e^{a_1 \ldots a_k}\eqdef e^{a_1}\wedge \ldots \wedge e^{a^k}$ and the
symbol $=_{U}$ means that equality holds only after restriction of $\omega$ to
$U$ . A bundle of real {\em pinors} over $M$ is an $\R$-vector bundle $S$ over $M$ 
which is (compatibly) a bundle of modules over the Clifford bundle $\Cl(T^\ast
M)$. Similarly, a bundle of real {\em spinors} is a bundle of modules over the even
Clifford bundle $\Cl^\ev(T^\ast M)$. Of course, a bundle of
real pinors is automatically a bundle of real spinors. Hence any pinor
is naturally a spinor but the converse need not hold. In this paper,
we focus on the case of {\em pinors} and in particular on the case when $S$ is
a bundle of {\em simple} modules over $\Cl(T^\ast M)$.  We let $\gamma^m\eqdef
\gamma(e^m)\in \Gamma(U,\End(S))$ and $\gamma_m\eqdef \eta_{mn}\gamma^n\in
\Gamma(U,\End(S))$  be the contravariant and covariant `gamma matrices' associated with the local orthonormal
coframe $e^m$ of $M$ and let $\gamma_{m_1\ldots m_k}$ denote the complete
antisymmetrization of the composition $\gamma_{m_1}\circ \ldots \circ
\gamma_{m_k}$. 

\paragraph{Twisted (anti-)selfdual forms.}
Assuming that $M$ is oriented, we let $\nu$ denote the volume form of
$(M,g)$. We concentrate on the case when $\nu\diamond \nu=+1$ (this
happens, in particular, when $M$ is an eight- or nine-manifold endowed with a
Riemannian metric --- the two cases which will be relevant for our
application). With this assumption, we have the bundle decomposition: 
\be
\wedge T^\ast M=(\wedge T^\ast M)^+ \oplus (\wedge T^\ast M)^-~~,
\ee
where the spaces of sections of the bundles $(\wedge T^\ast M)^\pm $ of
{\em twisted (anti-)selfdual forms} are the following $\cinf$-submodules of $\Omega(M)$: 
\be
\Omega^\pm(M)\eqdef \Gamma(M,(\wedge T^\ast M)^\pm)=\{\omega\in \Omega(M)|
\omega\diamond \nu=\pm \omega\}~~.
\ee
When $g$ is a Riemannian metric, the bundle morphism $\gamma:\wedge T^\ast
M\rightarrow \End(S)$ is injective iff. $d\not \equiv_8 1,5$. When $d\equiv_8
1,5$, we have $\gamma(\nu)=\epsilon_\gamma\id_S$, where $\epsilon_\gamma\in
\{-1,1\}$ is a sign factor known \cite{ga1} as the {\em signature of
  $\gamma$} . In those cases, we have $\gamma |_{(\wedge
  T^\ast M)^{-\epsilon_\gamma}}=0$ while the restriction $\gamma |_{(\wedge
  T^\ast M)^{+\epsilon_\gamma}}$  is injective. To uniformly treat all cases, we set: 
\be
(\wedge  T^\ast M)^\gamma\eqdef \twopartdef{\wedge  T^\ast M~~,~~}{d\not
  \equiv_8 1,5}{(\wedge   T^\ast M)^{+\epsilon_\gamma}~~,~~}{d \equiv_8 1,5}
\ee
and $\Omega^\gamma(M)\eqdef \Gamma(M,(T^\ast M)^\gamma)$, which is a
subalgebra of the \KA algebra. 

\paragraph{Dequantization.}
In our applications --- when $M$ is a Riemannian eight-manifold (the
compactification space of $M$-theory down to 3 dimensions) or a
nine-manifold (the metric cone over an eight-dimensional compactification space) ---
the fiberwise representation given by $\gamma$ is equivalent with
an irreducible representation of the real Clifford algebra $\Cl(8,0)$ or
$\Cl(9,0)$ in a 16-dimensional $\R$-vector space, which happens to be
surjective.  Due to this fact, the map $\gamma^{-1}\eqdef (\gamma|_{(\wedge T^\ast M)^\gamma})^{-1}:\End(S)\rightarrow
(\wedge T^\ast M)^\gamma$ can be used to identify the bundle of endomorphisms of $S$
with the bundle of algebras $((\wedge T^\ast M)^\gamma,\diamond)$. In particular, every globally-defined
endomorphism $T\in \Gamma(M,\End(S))$ admits a {\em dequantization}
$\check{T}\eqdef \gamma^{-1}(T)\in \Omega^\gamma (M)$, which
is a (generally inhomogeneous) differential form defined on $M$. Furthermore,
the dequantization of a composition $T_1\circ T_2$ equals the geometric product
$\check{T}_1\diamond \check{T_2}$ of the dequantizations of $T_1, T_2\in
\Gamma(M,\End(S))$. 

\paragraph{The Fierz isomorphism.}

When the Schur algebra  \cite{ga1} of $\Cl(p,q)$  is isomorphic with
$\R$ (i.e. when $\gamma$ is surjective), one can define an isomorphism of bundles of algebras
$\check{E}:(S\otimes S,\bullet) \stackrel{\sim}{\rightarrow} ((\wedge T^\ast
M)^\gamma, \diamond)$ called \emph{the Fierz isomorphism}, where $(S\otimes
S,\bullet)$ is a bundle of algebras known as the {\em bipinor
bundle}. On sections, this induces an isomorphism of $\cinf$-algebras
$\check{E}:(\Gamma(M,S\otimes S),\bullet) \stackrel{\sim}{\rightarrow}
(\Omega^\gamma (M), \diamond)$ which identifies the {\em bipinor algebra}
$(\Gamma(M,S\otimes S),\bullet)$ with the subalgebra $(\Omega^\gamma (M),\diamond)$ of
the \KA algebra. Both $\check{E}$ and the multiplication $\bullet$ of the
bipinor algebra depend on the choice of an {\em admissible} \cite{AC0, AC1} 
pairing $\cB$ on $S$.   In our application (when $M$ is an eight- or
nine-dimensional Riemannian manifold), $\cB$ is a certain admissible bilinear pairing on $S$
which is positive-definite and symmetric.  We define $\check{E}_{\xi,\xi'}\eqdef \check{E}(\xi\otimes
\xi')\in \Omega^\gamma(M)$, where $\xi,\xi'\in \Gamma(M,S)$.  

\paragraph{Constrained generalized Killing forms.} 

Using properties of the Fierz isomorphism, the algebraic constraint $Q\xi=0$
and the generalized Killing pinor equations $D\xi=0$ translate \cite{ga1}
into the following conditions on the inhomogeneous differential forms
$\check{E}_{\xi,\xi'}$, which hold for any global sections $\xi, \xi'\in
\Gamma(M,S)$ satisfying \eqref{par_eq}:
\ben
\begin{split}
\label{CGK}
\D \check{E}_{\xi,\xi'}=\check{Q}\diamond \check{E}_{\xi,\xi'}=0~~. 
\end{split}
\een
Here, $\check{Q}\eqdef \gamma^{-1}(Q)\in \Omega(M)$ is the `dequantization' of the globally-defined endomorphism $Q\in \Gamma(M,\End(S))$
and $\D=e^m\otimes \D_m$ is the `adjoint dequantization' of $D$ (see
\cite{ga1}).  The operators $\D_m$ are even derivations of the \KA algebra which are defined through:
\be
\begin{split}
\D_m\eqdef \nabla_m+[\check{A}_m,~]_{-,\diamond}~~,
\end{split}
\ee
where $\check{A}_m\eqdef \gamma^{-1}(A_m)$ and $\nabla$ is the connection
induced on $\wedge T^\ast M$ by the Levi-Civita connection of $(M,g)$.  The
Fierz identities between the form-valued pinor bilinears
$\check{E}_{\xi,\xi'}$ take the concise form \cite{ga1}:
\be
\begin{split}
\check{E}_{\xi_1,\xi_2}\diamond\check{E}_{\xi_3,\xi_4}=\cB(\xi_3,\xi_2)\check{E}_{\xi_1,\xi_4}~~,~~\forall
\xi_1,\xi_2,\xi_3,\xi_4\in \Gamma(M,S)~~,
\end{split}
\ee
defining a certain subalgebra of the \KA algebra of $(M,g)$.

Equations \eqref{CGK} generalize the usual theory of Killing forms in a number
of different directions and can be taken as a starting point for a
mathematical theory which is of interest in its own right. When expanding the
geometric product into generalized products as in \eqref{starprod}, these
seemingly innocuous equations become a highly non-trivial system whose
analysis would be extremely difficult without recourse to the synthetic
formulation given above in terms of \KA algebras. In particular, the geometric
algebra formulation given here allows one to easily extract structural
properties of such equations and to study them using techniques familiar from
the theory of non-commutative algebras and modules over such -- thereby
providing an interesting connection between spin geometry and noncommutative
algebraic geometry. We stress that equations \eqref{CGK} apply in much more
general situations than those considered in this brief summary.

\section{The CGK equations for metric cones}

As explained in \cite{ga2}, it is convenient to lift $\xi$ to the metric cone over $M$, which can be viewed as the warped product $({\hat M}, g_\cone)\approx
((0,\infty),\dd r^2) \times_r (M, g)$ (of warp factor equal to
$r$):
\be
\dd s_\cone^2=\dd r^2 +r^2 \dd s^2~~  .
\ee
The one-form
\be
\theta\eqdef \dd r= \partial_r \lrcorner g_\cone~~
\ee
has unit norm with respect to the cone metric. The pin bundle ${\hat S}$ of
the cone can be identified with the pullback of $S$ through the natural
projection $\Pi:{\hat M}\rightarrow M$. We define the {\em lift} ${\hat
  D}$ of $D$ to be the connection on ${\hat S}$ obtained from $D$ by pullback
to the cone.
Then ${\hat D}$ can be expressed as: 
\ben
\label{Dcone}
{\hat D}=\nabla^{{\hat S},\cone}+A^\cone~~,
\een
where $\nabla^{{\hat S},\cone}$ is the connection induced on ${\hat S}$ by the
Levi-Civita connection of $g_\cone$.  Since the metric cone $({\hat M},g_\cone)$ over $(M,g)$ has signature $(9,0)$
and since $9-0\equiv_8 1$, the Clifford algebra $\Cl(9,0)$ corresponds to the
normal non-simple case discussed in \cite{ga1}.  In particular, its Schur
algebra equals the base field $\R$ and the corresponding pin representation
$\gamma_\cone:(\wedge T^\ast {\hat M},\diamond^\cone)\rightarrow \End({\hat S})$ is surjective. 
We have two inequivalent choices for $\gamma_\cone$, which are distinguished by the signature $\epsilon\in \{-1,1\}$.  The morphism
$\gamma_\cone:(\wedge T^\ast {\hat M},\diamond^\cone) \rightarrow (\End({\hat
S}),\circ) $ is completely determined by the morphism $\gamma:(\wedge T^\ast
M, \diamond)\rightarrow (\End(S),\circ)$ once the signature $\epsilon$ has
been chosen. In the following, we shall work with the choice
$\epsilon=+1$. Setting $\epsilon=+1$ and rescaling
the metric on $M$ as $g\rightarrow (2\kappa)^2 g$, we find:
\ben
\label{scondplus1}
\nabla^{{\hat S},\cone}_{m} =\nabla^S_{m}+\kappa \gamma_{m9}~~,~~
A^\cone_9=0~~~,~~A^\cone_m=\frac{1}{4}f^p\gamma_{m p 9}+\frac{1}{24}F_{m p q r}\gamma^{p q r}~~.\nn
\een
The generalized Killing pinor equations $D_m\xi=0$ ($m=1\ldots 8$) for pinors
$\xi\in \Gamma(M,S)$ defined on $M$ amount to the flatness conditions:
\be
{\hat D}_a {\hat\xi}=0~~,~~\forall a=1\ldots 9~~,
\ee
 for pinors ${\hat \xi}\in \Gamma({\hat M},{\hat S})$ defined on ${\hat M}$. Indeed,
the last of the cone flatness equations ${\hat D}_9 {\hat \xi}=0$ is
equivalent with the requirement that the section ${\hat \xi}$ of ${\hat S}$ is
the pullback of some section $\xi$ of $S$ through the natural projection $\Pi$
from ${\hat M}$ to $M$, while the remaining equations amount to the
generalized Killing conditions $D_m\xi=0$ on $M$. Furthermore, the algebraic
constraint for $\xi$ is equivalent with the following equation for ${\hat \xi}$:
\be
{\hat Q} {\hat \xi}=0~~,
\ee
where ${\hat Q}\in \Gamma({\hat M},\End({\hat S}))$ is the pullback of
$Q\in \Gamma(M,\End(S))$. We refer the reader to \cite{ga2} for much more
detail about the geometric algebra realization of the cone formalism of
\cite{Bar} and for the applications of this realization to the theory of
constrained generalized Killing pinors and forms. 

\section{Application to $N=2$ compactifications of M-theory down to three dimensions}

In this example, one obtains useful simplifications of the problem by using
the geometric algebra reformulation (see \cite{ga2} and the previous Section) of the cone formalism, which is
particularly relevant when seeking a geometric interpretation in terms of
reductions of structure group. Using this variant of the cone formalism as
well as a software implementation of our approach using {\tt Ricci} \cite{Ricci} and {\tt
Cadabra} \cite{Cadabra}, one can extract and analyze the cone reformulation of
\eqref{par_eq}. Since the detailed theory of the \KA algebra of cones is
somewhat involved and since the equations obtained in this manner for the
application at hand are quite complex, we cannot reproduce them here given the
space limitations.  Instead, we refer the interested reader to \cite{ga2} and
\cite{mirela}.

\section{Conclusions}

We summarized an approach to the theory of constrained generalized Killing
(s)pinors which is inspired by geometric algebra, a formulation of spin geometry
which resolves the lack of naturality affecting certain traditional
constructions. Using this approach, we showed how generalized Killing pinor
equations translate succinctly into conditions for differential forms
constructed as bilinears in such pinors. We also touched upon the applications of this
approach to the study of $N=2$ compactifications of $M$-theory down to three
dimensions, which are discussed in more detail in \cite{ga2} as well as in \cite{mirela}.

\begin{acknowledgement}
This work was supported by the CNCS projects PN-II-RU-TE (contract number
77/2010) and PN-II-ID-PCE (contract numbers 50/2011 and 121/2011). The authors
thank the organizers of the 8-th QFTHS Conference for hospitality and interest
in their work.  C.I.L thanks the Center for Geometry and Physics, Institute
for Basic Science and Pohang University of Science and Technology (POSTECH),
Pohang, Korea for providing excellent conditions at various stages during the
preparation of this work, through the research visitor program affiliated with
Grant No. CA1205-1. The Center for Geometry and Physics is supported by the
Government of Korea through the Research Center Program of IBS (Institute for
Basic Science).  He also thanks Perimeter Institute for hospitality for
providing an excellent and stimulating research environment during the last
stages of this project. Research at Perimeter Institute is
supported by the Government of Canada through Industry Canada and by the
Province of Ontario through the Ministry of Economic Development and
Innovation.
\end{acknowledgement}

\end{document}